# On conservation of the of crystal lattice symmetry in transition at Curie point in exchange magnets

Khisa Sh. Borlakov[*] and Albert Kh. Borlakov

*North Caucasian State Humanitarian and Technological Academy,*

*36 Stavropolskaya str., Cherkessk, Russia, 369001*

We show that symmetry of the crystal lattice of exchange magnets (containing only 3d magneto-active elements) does not change at the Curie point; only the magnetic symmetry of the crystal is decreasing in the transition point. In the non-exchange magnets (containing only rare-earth magneto-active elements), on the contrary, both the magnetic and crystal-chemical symmetry decrease at the Curie point. There is isotropic magnetic phase in exchange magnets; and their magnetic symmetry is described by color groups of magnetic symmetry of P-type. Non-exchange magnets do not have isotropic phase; their symmetry is described by color groups of magnetic symmetry of Q-type.

## 1. Introduction

Operations of space symmetry groups reduce to translation of an atom with a coordinate $\vec{r}$ into a crystallographically equivalent position with a coordinate $\vec{r}'$. Symbolically, this can be written as follows: $g(\vec{r}) = \vec{r}'$. If the atoms possess some physical characteristics, then translation of an atom may lead to a change in the corresponding physical characteristic. The conventional space groups describe symmetry properties of crystal lattices occupied by structureless atoms. If the space lattice is occupied by atoms with physical characteristics, then the description of its symmetry properties requires color symmetry groups to be used. Let the additional physical characteristic possessed by an atom be a spin vector $\vec{S}$, which we will represent by a classical vector. Then the symmetry operation can be represented in the following form:

$$\{g(\vec{r}) \cdot A(\vec{S})\} = \{\vec{r}', \vec{S}'\}. \qquad (1)$$

This generalization of the concept of space (Fedorov's) groups is called color (magnetic) groups of the $P$ type [1]. Noncrystallographic operations $A$ entering into (1) are called operations of loads, whereas the operations of the space group are called operations of the basis [1]. As can be seen from (1), the operations of the basis do not change the spin direction.

The color magnetic group can also be defined in another way using the relationship

$$\{g(\vec{r}, \vec{S}) \cdot A(\vec{S})\} = \{\vec{r}', \vec{S}'\}. \qquad (2)$$

In definition (2), the operation of the basis $g$ changes not only the atom position, but also its spin direction. The color symmetry group defined in this way is called a $Q$- type group [1]. For example, the conventional Shubnikov groups of magnetic symmetry are color magnetic groups of the $Q$- type with a single load $A = 1'$, i.e., the operation of spin inversion (anti-identification in Shubnikov's notation). There also exist other ways to generalize classical Fedorov groups (see [1,2]) and references therein), but for our purposes it will suffice to consider the above two types of color magnetic symmetry.

For each color magnetic group, there exists a maximal space subgroup $G$, which is obtained from the first if all the loads are taken to be identical elements; this can be written symbolically as

$$\lim_{\{A_i\} \to 1} M = G \qquad (3)$$

Relationship (3) was called the Landau-Lifshitz criterion in [1], and a certain theoretical status was given to it. If (3) is understood as denoting that each magnetic group has a space subgroup that coincides with the symmetry group of the crystal-chemical lattice, then this is a very trivial statement, which does not deserve to have theoretical status. However, the situation is not so simple and Izyumov *et al.* [1] actually had grounds to believe that (3) is not a trivial group-theoretical relation.

Let us consider this problem in more detail, following [3].

## 2. Results and discussion

The Landau-Lifshitz criterion becomes a physically meaningful statement if we interpret it in the following way: There exist magnetically ordered phases whose space group coincides with the space group of the corresponding paramagnetic phase. Are there some experimental facts that correspond to this statement, and what does this mean if such facts exist? As is well known, a decrease in the symmetry of a crystal lattice upon transition from a paramagnetic to a magnetically ordered phase is due to relativistic interactions [4]. The relativistic interactions ensure interaction between the magnon and phonon subsystems. This is manifested in magnetostriction strains of the lattice and in the appearance of axes of easy and hard magnetization (magnetic anisotropy). It is commonly accepted that magnetic anisotropy arises at the Curie-Neel point. This is indeed observed in rare-earth metals and their magnetic compounds. As for *3d* elements and their magnetic compounds, the picture is quite different. For example, for $\alpha$-iron the curve of the temperature dependence of spontaneous magnetization terminates (as the temperature increases from zero) at the Curie point $T_c = 1043$ K, whereas the curves of the temperature dependences of the magnetic anisotropy constants terminate at a lower temperature (see figure). As can be seen from the figure, no magnetic anisotropy exists in iron in the temperature range of 950-1043 K.

This means that in this range a magnetically isotropic phase exists in iron whose magnetic properties are only caused by exchange interaction. In this case, the temperature $T_{ls} \approx 950$ $K$ must represent the critical temperature of a phase transition from an isotropic to an anisotropic phase and a $\lambda$-type thermal anomaly must exist at this temperature. Such an anomaly was actually revealed in [6]. Quite similar behavior of magnetic properties is characteristic of nickel, cobalt, and some magnetic compounds of the elements of the iron group [7-9], Thus, in the *3d* elements and their magnetic compounds, there exists an isotropic magnetic phase and a relativistic phase transition from an isotropic to an anisotropic magnetic phase.

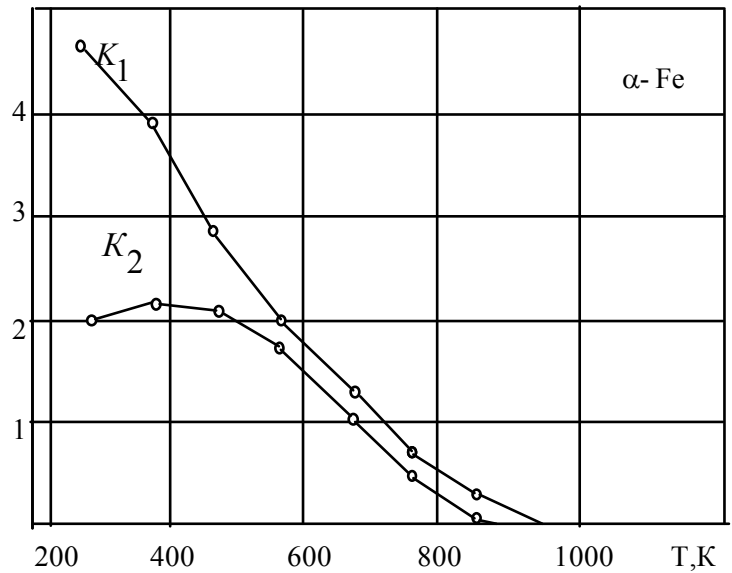

**Figure**. Temperature dependence of the magnetic anisotropy constants of the $\alpha - Fe$

Koptsik noted (see [10] and references therein) that color magnetic groups of the *P* type describe the magnetic symmetry of crystals without allowance for relativistic interactions, whereas the color groups of the *Q* type describe magnetic symmetry with allowance for relativistic interactions. Most authors are sure that it does not matter whether or not we take relativistic interactions into account in calculations of magnetic structures. The above facts suggest the opposite. As the temperature in an iron crystal decreases from 1043 K, a spontaneous ferromagnetic moment arises in it, which is oriented in an arbitrary way with respect to the crystal axes. This arbitrariness of the magnetization orientation exists down to a temperature $T_{ls}$ of about 950 K, below which the magnetization becomes oriented along one of the cube edges and the bcc unit cell becomes tetragonally distorted, with the tetragonal axis coinciding with this preferred edge. Thus, Nature takes no interest in whether or not we take into account relativistic interactions and acts in its own way. With the above in mind, we may interpret Koptsik's conclusion in the following way: Color magnetic groups of the *P* type quite exactly describe the magnetic symmetry of isotropic phases, whereas a description of the symmetry of anisotropic phases requires resorting to magnetic groups of the *Q* type. The differences between anisotropic and isotropic phases and between color magnetic groups of the *P* and *Q* types can be interpreted mechanistically. Relationship (1) determines the result of the action of a symmetry operation in a color magnetic group of the *P* type. It can be seen from this relationship that the operation of changing the atom's position and the operation of changing the direction of its spin are independent of each other. This may be only caused by the fact that there is no relation between the spin subsystem and the lattice in reality. In an anisotropic phase such a relation appears, relationship (1) is no longer valid, and we should use formula (2). Let there be a gyroscope in gimbals at each magnetic site of the lattice. Lattice rotation will not change the orientations of the rotation axes of the gyroscopes relative to the laboratory coordinate system; and *vice versa,* we may change the directions of rotation axes of the gyroscopes without rotating the lattice. This is a mechanistic model of an isotropic phase. Now, let the gyroscope axes be rigidly fixed to the lattice, so that any change in the orientation of the lattice in space will change the orientation of the rotation axes of the gyroscopes relative to the laboratory coordinate system. The orientation of the rotation axes relative to the coordinate axes rigidly fixed to the lattice does not, however, change in this case. This is a mechanistic model of an anisotropic phase or nonexchange magnet.

Since, in the isotropic state, the crystal lattice of a magnet coincides exactly with the lattice in the paramagnetic state, the space groups of the paramagnetic and isotropic phases coincide and the Landau-Lifshitz criterion is satisfied just in its physically meaningful variant. For example, the symmetry group of the paramagnetic phase of a iron is $O_h^9 \times O(3)$ and the symmetry group of the isotropic ferromagnetic phase is $O_h^9 \times O(1)$. Here, O(3) is the three-parameter group of spin rotations and O(1) is the single-parameter group of spin rotations whose axis coincides with the magnetization vector. Evidently, the space subgroups of both these groups coincide with one another in full agreement with the Landau-Lifshitz criterion. As for the anisotropic ferromagnetic phase with its $Q$ type symmetry group $I4/mm'm'$, the Landau-Lifshitz criterion is not fulfilled in this case.

Thus, we argue that the symmetry of a crystal lattice always decreases only below the temperature of the relativistic transition (for exchange magnets). The uniaxial crystals of the rhombohedral, hexagonal, and tetragonal systems seem to be exceptions. We might expect that, below the point of transition to the anisotropic phase, no reduction in the symmetry of the crystal lattice would occur in the easy-axis magnets; moreover, many authors are firmly convinced that this is the case. However, this is not the case in reality. The irreducible representations that are related to the ferromagnetic ordering of the easy-axis type for point groups of uniaxial crystals are as follows [5,11]:

$D_4(A_2); C_{4v}(A_2); D_{4h}(A_{2g}); D_3(A_2); C_{3v}(A_2); D_{3d}(A_{2g}); D_{3h}(A_2); D_6(A_2); D_{6v}(A_2); D_{6h}(A_{2g})$.

It is well known from the Landau theory of phase transitions [12,13] that only identity irreducible representations induce phase transitions without a change in the symmetry of the crystal lattice. But for the above-listed uniaxial point groups, only the $A_1$ and $A_{1g}$ irreducible representations are identical [14]. Therefore, even in the ferromagnetic state of the easy-axis type the symmetry of the crystal lattice of uniaxial crystals is lower than in the paramagnetic state. Thus, the Landau-Lifshitz criterion is not fulfilled in anisotropic magnetic phases even in the case of uniaxial crystals.

## 3. Conclusion

For P-type color magnetic groups that describe the magnetic symmetry of isotropic phases, a Landau-Lif-shitz criterion is fulfilled, which states that the space symmetry group of a magnetically ordered crystal coincides with the space group of the paramagnetic phase. This statement has nothing in common with the trivial affirmation that the space group of a crystal is a subgroup of its magnetic symmetry group, which can be deduced superficially from (3). If the isotropic magnetic phase did not exist, the Landau-Lifshitz criterion would not be fulfilled for any of the magnets.

---

[*] Electronic address: borlakov@mail.ru